\documentclass[aps,notitlepage,nofootinbib,longbibliography,amsmath,amssymb,
twocolumn,
superscriptaddress,10pt]{revtex4-1}

\usepackage[usenames,dvipsnames]{xcolor}
\definecolor{bluepurple2}{rgb}{0.06,0,0.6}
\usepackage[colorlinks=true,citecolor=blue,linkcolor=bluepurple2]{hyperref}

    \usepackage{amsmath}
\usepackage{amsthm, amssymb}
\usepackage{enumitem}
    \usepackage{graphicx}
    \usepackage{color}
    \usepackage{bbold}

    \newcommand{\bit}{\begin{itemize}}
    \newcommand{\eit}{\end{itemize}}

    \newcommand{\f}{\frac}
    \renewcommand{\>}{\right\rangle}
    \newcommand{\<}{\left\langle}
    \newcommand{\ba}{\begin{align}}
    \newcommand{\ea}{\end{align}}
    \newcommand{\be}{\begin{equation}}
    \newcommand{\ee}{\end{equation}}
    \newcommand{\bi}{\begin{itemize}}
    \newcommand{\ei}{\end{itemize}}
    \newcommand{\lf}{\left(}
    \newcommand{\ri}{\right)}
    \newcommand{\dd}{\mathrm{d}}

    \def\+{\dagger}

    \newcommand{\newsec}[1]{\emph{ #1}}

\begin{document}

    \newcommand{\bra}[1]{\< #1 \right|}
    \newcommand{\ket}[1]{\left| #1 \>}

        \newcommand{\bbra}[1]{\<\< #1 \right| \right|}
    \newcommand{\kket}[1]{\left|\left| #1 \>\>}
    
           \newcommand{\bbraket}[2]{\<\< #1 || #2 \>\>}
    
    \title{Fixed point annihilation for a spin in a fluctuating field}
    \author{Adam Nahum}
\affiliation{Laboratoire de Physique, \'Ecole Normale Sup\'erieure, CNRS, Universit\'e PSL, Sorbonne Universit\'e, Universit\'e de Paris, 75005 Paris, France.}

\begin{abstract}
A quantum spin impurity coupled to a critical free field 
(the Bose-Kondo model)
can be represented as a 0+1D field theory with long-range-in-time interactions that decay as $|t-t'|^{-(2-\delta)}$. This theory is a simpler analogue of nonlinear sigma models with topological  Wess-Zumino-Witten terms in higher dimensions. 
In this note we show that the RG flows for the impurity problem exhibit an annihilation between two nontrivial RG fixed points at a critical value $\delta_c$ of the interaction exponent.
The calculation is controlled at large spin $S$.
This  clarifies the phase diagram of the Bose-Kondo model and shows that it serves as a  toy model for phenomena involving fixed-point annihilation and ``quasiuniversality'' in  higher dimensions.
\end{abstract}

    \date{\today}
  \maketitle

The annihilation of a stable with an unstable fixed point is a generic possibility in renomalization group (RG) flows when a parameter such as the spatial dimensionality, which does not flow, is varied
\cite{nienhuispotts,cardynauenbergscalapino,newman1984q,zumbach1993almost,Gies,kaplan,GukovRG2017}.  
When this happens it leads to an interesting regime just beyond the annihilation point. No physical fixed point exists in this regime (though ``annihilation'' really means that the real fixed points disappear into the complex plane, where they may correspond to nonunitary conformal field theories \cite{gorbenko2018walking}).
Nevertheless the RG flows become very slow. 
This can yield particles with anomalously small masses, 
 or weakly first-order phase transitions with extremely long correlation lengths \cite{nienhuispotts} that show quasiuniversal \cite{zumbach1993almost,wang2017deconfined} behaviour below this scale.

One generic class of examples includes field theories with cubic terms that have continuous transitions in low dimensions, which become first order (as predicted by mean-field theory) in high dimensions. These include the Potts model \cite{newman1984q}
(which also undergoes annilation in 2D as a function of the number of states \cite{nienhuispotts,cardynauenbergscalapino,gorbenko2018walking2,iino2019detecting,ma2019shadow})
as well as Landau theories for order parameters on complex or real projective space \cite{PhaseTransitionsCPNSigmaModel, sernanahumforthcoming}.

This note is motivated by a fixed point annihilation  phenomenon that was proposed to resolve debates about Monte Carlo results for deconfined criticality \cite{deccp} in 2+1D antiferromagnets \cite{DCPscalingviolations,wang2017deconfined}.  In Refs.~\cite{ma2020theory,nahum2020note} this was put in terms of a dimensional hierarchy of nonlinear sigma models in $d$ spacetime dimensions with  ${\mathrm{SO}(d+2)}$ global symmetry \cite{AbanovWiegmann}. These sigma models have a topological Wess-Zumino-Witten (WZW)  term in the action. The case ${d=2}$ is the well-known WZW theory with a conformal fixed point \cite{witten1984non}, and ${d=3}$ is an effective field theory for various competing order parameters in {2+1D} magnets \cite{tanakahu,tsmpaf06}.  It was argued that fixed point annihilation occurs between two and three dimensions.

Unfortunately, this example of fixed point annihilation, like the others mentioned above, requires an integer-valued parameter (here $d$) to be treated as  continuously variable.  
An annihilation that takes place at a noninteger critical dimensionality may be useful conceptually for understanding nearby values of $d$, 
but it cannot be realized physically (and there may be ambiguities in defining the continuous $d$ theory).
It would be instructive to have a toy model  that retained basic features of the WZW example, without the unphysical feature of noninteger $d$.

\begin{figure}[b]
\centering
\includegraphics[width=0.9\columnwidth]{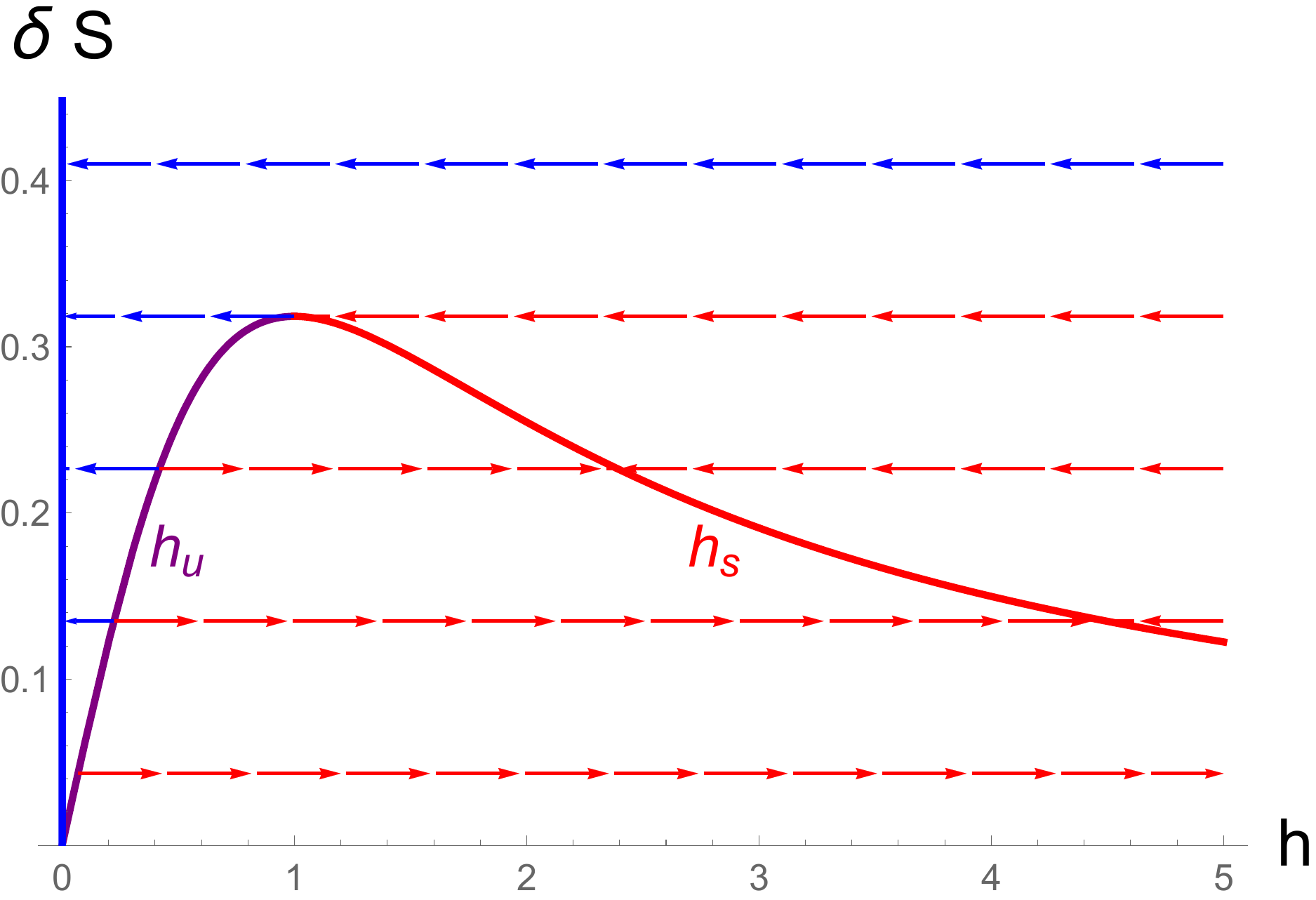}
\caption{
Fixed points and flows as a function of the exponent $\delta$ in the memory kernel $K\sim 1/|t-t'|^{2-\delta}$. 
$h$ is the dimensionless coupling of the 1D nonlinear sigma model.  $h=0$ is the ordered fixed point and $h=\infty$ (not shown) is a noninteracting spin with $2S+1$ degenerate ground states. $h_{s,u}= S g_{s,u}$ label the branches of stable and unstable fixed points.}
\label{fig:flows}
\end{figure}

Here we show that the simplest member of the ``WZW'' hierarchy, in $d=1$, provides such a model if we augment it with a long-ranged interaction. 
This is a model of a spin impurity in a gapless environment \cite{sengupta2000spin,smith1999non,sachdev1993gapless,vojta2006impurity,sachdev2004quantum}, and was suggested as a model for fixed point annihilation in  \cite{nahum2020note}. 
We find that many key features of the higher-dimensional example are retained  (fixed point annihilation, quasiuniversality, emergent symmetry).
But since the fixed point annihilation occurs in $d=1$ the model is accessible to numerical simulations and perhaps to experiment.
The model is analytically tractable at large spin.

The $d=1$ theory without a long-range interaction is simply the quantum mechanics of a spin-$S$ (or more generally a rotor), described using the spin path integral with its well-known Berry phase term \cite{altland2010condensed}.
The version with a power-law interaction ${\sim 1/|t-t'|^{2-\delta}}$ describes a spin with a retarded interaction, physically representing an interaction with a  gapless zero-temperature bath that has been integrated out.
This is known as the ``Bose-Kondo'' model \cite{sengupta2000spin,smith1999non,sachdev1993gapless,vojta2006impurity,sachdev2004quantum,sachdev1999quantum,vojta2000quantum,si2001locally,zarand2002quantum,zhu2002critical,vojta2003pseudogap,zhu2004quantum,novais2005frustration,pixley2013quantum,chowdhury2021sachdev,joshi2020deconfined}. It falls into the larger family of quantum impurity models describing a local quantum-mechanical degree of freedom interacting with a bath of critical fluctuations \cite{vojta2006impurity,leggett1987dynamics,sachdev2004quantum,hewson1997kondo,allais2014spectral}.

We study the model in a large spin limit that allows the RG equation to be obtained to all orders in the coupling. 
Using the background field method, the calculation is a simple extension (to include the Berry phase term) 
of the analysis by Kosterlitz of the long-range classical Heisenberg model  in one dimension \cite{kosterlitz1976phase}. 
The beta function shows an interesting structure, with two nontrivial fixed points that annihilate with each other when the interaction power law  ${2-\delta}$ is varied through a critical value. The flows as a function of $\delta$ are qualitatively like those suggested for the WZW model as a function of $\epsilon$ (in $2+\epsilon$ dimensions) \cite{ma2020theory,nahum2020note}, except in the behaviour of one of the nontrivial fixed points (the stable one) when~${\delta\rightarrow 0}$.

\newsec{Model.} We consider a Euclidean action for a spin of size $S$ with a long-ranged temporal interaction:
\ba\label{eq:action1}
\mathcal{S} = \f{1}{2g} \int \dd t \dd t'\, K(t-t') \big( \vec n(t) - \vec n(t') \big)^2  - i S \,\Omega[\vec n].
\end{align}
Here ${\vec n=(n_1,n_2,n_3)}$, with ${\vec n^2=1}$, is the field appearing in the coherent states path integral.
This is a formulation of the $\mathrm{SO}(3)$-symmetric Bose-Kondo model,
in which the spin is coupled  to a local magnetization ${\vec m}$
(associated with additional ``bulk'' degrees of freedom) 
via a Hamiltonian ${H_\text{int} = J {\vec S}.{ \vec m}}$ \cite{sengupta2000spin,smith1999non,sachdev1993gapless,vojta2006impurity,sachdev2004quantum}.
If $\vec m$ has $\mathrm{SO}(3)$-invariant autocorrelations obeying Wick's theorem, then integrating ${\vec m}$ out yields (\ref{eq:action1}) with ${g^{-1}\propto J^2 S^2}$ and with a kernel ${K(t-t')}$ that is proportional to the autocorrelator of ${\vec m}$. We assume this to be a power law at large ${\tau=t-t'}$, ${K(t-t')\propto |t-t'|^{-(2-\delta)}}$ with  ${-1<\delta<1}$. For convenience we normalize $K$ as 
\ba\label{eq:Kdefn}
K(\tau) & = \f{C \Lambda^\delta}{|\tau|^{2-\delta}}, & 
C & =\f{(1-\delta)}{4\Gamma(\delta)\sin(\pi\delta/2)}.
\end{align}
The constant $C$ is chosen so that the Fourier transform of $K(\tau)$ has a simple normalization (App.~\ref{app:calculation}) \cite{kosterlitz1976phase},
and a power of the UV cutoff frequency $\Lambda$  is included in $K(\tau)$ so that $g$ is dimensionless.
Finally, the Berry phase term $\Omega[\vec n]$ is the solid angle on the sphere traced out by the trajectory, written in terms of an extension of the field $\vec n(t)$  to a field $\vec n(t,u)$ defined on a strip with $u\in[0,1]$ as~\cite{altland2010condensed}:
\ba
\Omega[\vec n] & =\f{1}{2} \int_0^1 \dd u \int \dd t  \epsilon^{\mu\nu} \vec n. (\partial_\mu \vec n\times \partial_\nu \vec n) 
\end{align}
or more simply as ${\Omega[\vec n] =\int \dd t (1-\sin\psi)\dot\phi}$  in the coordinates ${\vec n = (\cos \psi \cos \phi, \cos \psi \sin \phi, \sin\psi)}$.

Before calculating the beta function, let us ask what we can expect from stability considerations. 

The action (\ref{eq:action1}) has two trivial fixed points, at $g=0$ and at $g=\infty$. That at $g=0$ is a perfectly ordered state, with no local fluctuations in $\vec n(t)$.
The fixed point at $g=\infty$ describes a quantum spin with ${2S+1}$ degenerate ground states.

By counting dimensions we see that when $\delta$ is negative the ordered fixed point at ${g=0}$ is unstable and the ${g=\infty}$ fixed point is stable. The simplest expectation (confirmed in the large $S$ calculation  below) is that in this ${\delta<0}$ regime the model flows, for any positive $g$, to ${g=\infty}$. On the other hand when $\delta$ is positive the ordered fixed point becomes stable, so the model is in a stable ordered phase for small enough $g$. At the same time the ${g=\infty}$ fixed point becomes unstable.

The flows for \textit{infinitesimal} $g$ are similar to those in a classical 1D model without the Berry phase term, because the Berry phase term in the action is subleading in the limit $g\rightarrow 0$. As in the classical model, the change in stability of the ordered fixed point is accompanied by the appearance of a nontrivial unstable fixed point, representing a phase transition, at a coupling $g_u$ that is of order $\delta$ for small positive $\delta$ \cite{kosterlitz1976phase}.

However  the Berry phase term plays a role for non-infinitesimal $g$. 
In particular, the  $g=\infty$ fixed point is unstable for $\delta>0$, 
unlike a simple classical disordered fixed point.
The simplest consistent hypothesis is therefore that for sufficiently small positive $\delta$ there is another nontrivial fixed point $g_s$, with $g_s> g_u$, which is \textit{stable}.   This fixed point governs a stable large-$g$ phase with power-law correlations.
[Heuristically, the Berry phase term has prevented $\vec n(t)$ from being trivially disordered at large $g$, leading instead to a stable ``critical phase''.] 
At small $\delta$, with $S$ fixed, this stable fixed point can be studied by perturbative RG in the strength of the impurity-spin coupling~\cite{vojta2006impurity}. 

What happens to these fixed points as $\delta$ is increased?  A simple guess (in analogy to the higher-dimensional problem) is that at some critical value $\delta_c$ they merge and annihilate, meaning that for a sufficiently long range interaction the model is always in the ordered phase (Fig.~\ref{fig:flows}). We will confirm this directly when $S$ is large. 

\newsec{RG results.} At large $S$ the interesting regime is where the coupling $g$ and the exponent $\delta$ are both of order $1/S$, so we will write
\ba
h & = g \,S , &
\widetilde \delta &=  \delta  \, S.
\end{align}
This scaling of the coupling ensures that the two terms in the action are of comparable size in the limit of large $S$. 
(If this is not the case, then one of the two terms dominates the action for the ``fast'' modes that we integrate out in the RG step, leading to a more trivial RG equation.)
The spin size $S$ itself is quantised and does not flow, but it serves as a large parameter that justifies a one-loop calculation \cite{witten1984non}. This calculation can be done with the background field method \cite{kosterlitz1976phase,polyakov1975interaction}  and is described in App.~\ref{app:calculation}.

Our basic result is the RG equation
\be\label{eq:basicrgeqn}
\f{\dd h}{\dd \tau} = 
\f{1}{S}\lf 
- \widetilde \delta \, h + \f{2}{\pi} \f{h^2}{1+h^2} \ri
+ \mathcal{O}\lf \f{1}{S^2} \ri,
\ee
where the RG time $\tau$ is the logarithm of a physical timescale.
The topology of the associated flows is shown in Fig.~\ref{fig:flows}.

The value 
\be
\delta_c = \f{1}{\pi S}
\ee
for the interaction exponent separates two regimes. For larger $\delta$, all flows lead to the ordered phase as noted above, but for ${0<\delta<\delta_c}$ there is stable nontrivial phase (governed by the fixed point at $g_s$), separated by a second-order phase transition (governed by $g_u$) from the ordered phase. The RG eigenvalue of the coupling at $g_{u,s}$ is ${y_g=\pm \delta\sqrt{1-\pi^2 \widetilde \delta^2}}$, with the $+$ sign for $g_u$.

The scaling dimension of the field $\vec n$ is $x_{1}=\delta/2$ at both  nontrivial fixed points, so that the spin autocorrelator decays as ${|t-t'|^{-\delta}}$. This exponent value is expected to be exact, as for other long range models, since the two local operators appearing in the long-range term renormalize independently when ${t-t'}$ is large \cite{fisher1972critical,kosterlitz1976phase,brezin1976critical,paulos2016conformal,zhu2002critical}. 
Below we will also need the scaling dimension $x_k$ of the symmetric $k$-index tensor $X^{(k)}_{a_1,\ldots,a_k}$ that is obtained as the traceless part of the  operator $n_{a_1}\ldots n_{a_k}$. At one-loop order this obeys \cite{brezin1976renormalization}  (App.~\ref{app:calculation})
\be\label{eq:xk}
x_k =  \f{k(k+1)}{2} x_1.
\ee

We conjecture that the topology of the flows found here at large $S$ applies for all values of the spin, including ${S=1/2}$. It would be interesting to study this numerically. 
The partition function for the spin has a Monte-Carlo-sign-free diagrammatic formulation, with propagators of $\vec m$ represented as arcs connecting points $t$, $t'$ on the spin's worldline \cite{weber2021quantum}, and the model may also be studied with numerical RG \cite{bulla2005numerical,glossop2005numerical}.

Let us return to the analogy with higher-dimensional models for deconfined criticality and competing orders.
In the  WZW hierarchy, two key features are 
 (1) quasiuniversality in the regime just beyond the fixed point annihilation ($\epsilon\gtrsim\epsilon_c$ in $2+\epsilon$ dimensions);
 and  (2) the emergence of the full  symmetry of the sigma model from a smaller microscopic symmetry group, thanks to  the irrelevance of operators analogous to $X^{(k)}$ for large enough $k$.
We examine analogues of these phenomena in the present system.

\newsec{Quasiuniversality.} The quasiuniversality phenomenon  will occur in this 1D model when $\delta\gtrsim \delta_c$.
The spin will ultimately be ordered even if the bare coupling $h$ is large,
 but this will not be apparent until a timescale $\xi$ that diverges \textit{exponentially} with ${(\delta-\delta_c)^{-1/2}}$, because  the flows spend a large amount of RG time close to ${h=1}$~\cite{nienhuispotts,cardynauenbergscalapino}.

At small ${\delta-\delta_c}$ we can continue to classify operators as relevant or irrelevant, and the long RG time spent close to $h=1$ means that irrelevant perturbations, which will be present in a generic microscopic model with the appropriate symmetry, become exponentially small in ${(\delta-\delta_c)^{-1/2}}$ \cite{wang2017deconfined}. 
This exponential suppression of differences between bare models underlies quasiuniversality. For example, we will have approximate universality in the functional form of the spin autocorrelator ${\<\vec n(t).\vec n(0)\>}$, despite the fact that it is not a power law for ${\delta>\delta_c}$. 

In fact, a simplifying feature of RG for the long-range model is that
 the  flow of the renormalized coupling $h(\tau)$ --- obtained from running the RG up to  a physical timescale
  ${\Lambda^{-1}e^\tau}$ ---
 can be plotted simply by plotting the spin autocorrelator, at least within the present large $S$ approximation.
This is because the RG equation  (\ref{eq:basicrgeqn}) can be expressed in terms of the running scaling dimension $x(h)$ of the sigma model field $\vec n$, as ${\dot h = (-\delta + 2 x(h)) h}$ (see App.~\ref{app:calculation},  Eq.~\ref{eq:x1h}). 
RG for the correlator then gives:
\be\label{eq:correlator}
\<\vec n(t).\vec n(0)\>  = \f{h(0)}{(\Lambda t)^\delta \, h(\ln \Lambda t)}.
\ee
It would be interesting to use the correlator (\ref{eq:correlator}) to obtain a proxy for the   beta function from Monte Carlo simulations in the quasiuniversal regime.

As an aside, a curious feature of the model is that the two point function (\ref{eq:correlator}) tends to a constant at large times both in the ordered (${h\rightarrow 0}$) phase, which is stable  for ${\delta >0}$,
and also in the free spin (${h\rightarrow \infty}$) phase that exists for $\delta<0$.
However the two fixed points are different. 

One concrete way to see the difference is in connected two-point functions $G^{(k)}(t)$ of  operators with higher spin,  ${k> 2S}$.   For a completely free spin, nonvanishing operators only exist with spin ${k\leq 2S}$. However, in the microscopic theory of a spin coupled to a bath we can construct nonvanishing operators with any spin $k$, as discussed in App.~\ref{app:calculation}.  Let $G^{(k)}(t)$ be a connected 2-point function for such operators. 
In the free spin phase $\lim_{t\rightarrow\infty} G^{(k)}(t)$ is nonzero for $k \leq 2S$ and vanishes for $k>2S$ 
(because the spin and bath decouple at the governing IR fixed point, see App.~\ref{app:calculation}). 
In contrast, in the ordered phase we expect that  ${\lim_{t\rightarrow\infty} G^{(k)}(t)}$  is nonzero for all $k$,
because the corresponding continuum operators $X^{(k)}$, defined above, have nonvanishing  long-distance correlations at the ordered fixed point. (Correlation functions at the ordered fixed point are simple since only the zero mode of $\vec n(t)$ needs to be averaged over.)

\newsec{Emergent symmetry.}  We can construct a simple toy model for the emergent symmetries [$\mathrm{SO}(4)$ in 1+1D or $\mathrm{SO}(5)$ in 2+1D] that arise in various higher-dimensional microscopic models for which the WZW sigma models serve as effective field theories~\cite{tanakahu,tsmpaf06,fradkin2013field,emergentso5,wang2017deconfined,sreejith2018emergent,ma2019role,tsvelik2007quantum,patil2018numerical}.  In these examples, the $N$-component sigma model field ${\vec n}$ is viewed as the concatenation of two separate fields,   ${\vec n = (\vec \Phi^{A}, \vec\Phi^{B})}$.
$\vec \Phi^{A}$ and $\vec \Phi^{B}$ are not related by microscopic symmetry, but may be related by an emergent $SO(N)$ symmetry at a critical point.
In 2+1D, for example, $\vec \Phi^{A}$ and $\vec \Phi^B$ could be the N\'eel and VBS order parameters, with the critical point of interest separating N\'eel and VBS phases.

Here we take ${\Phi^{A} = n_z}$ and ${\vec\Phi^{B} =  (n_x,n_y)}$.
That is, we think of Eq.~\ref{eq:action1} as an effective field theory for a phase transition in an \emph{anisotropic}  microscopic Hamiltonian, with only ${\mathrm{O}(2)= \mathbb{Z}_2 \ltimes \mathrm{U}(1)}$ symmetry, which is promoted to emergent ${\mathrm{SO}(3)}$ at a critical point.
The critical point lies at the boundary of a phase with easy-axis order for~$n_z$. 

For concreteness, consider simple ${\mathrm{O}(2)}$-invariant Hamiltonians for spin-1/2 and spin-1. Nontrivial examples  require at least two anisotropic couplings in the microscopic Hamiltonian, as will be clear below. 
For a spin-1 we could consider single-ion anisotropy and an anisotropic bath coupling:
 ${H_\text{aniso}= J \lf  S_x m_x + S_y m_y + \gamma S_z m_z \ri- \Delta S_z^2 }$.
 For a spin-1/2 the $S_z^2$ term trivializes, but we could consider a local anisotropy for the bath,
  ${H_\text{aniso}'= J \lf  S_x m_x + S_y m_y + \gamma S_z m_z \ri- \Delta m_z^2 }$.
 We assume that $\delta\leq \delta_c$, and that $J$ is small enough that the isotropic  models (${\Delta=0, \gamma=1}$)  flow to the fixed point at $g_s$, which is stable in the absence of anisotropy (Fig.~\ref{fig:flows}). 
 
Microscopic $\mathrm{O}(2)$ symmetry allows the perturbations ${\delta \mathcal{S} = \int \dd t \big( u X^{(2)}_{33} + v X^{(4)}_{3333}+\ldots \big)}$ to the continuum action.
Here ${X^{(2)}_{33}\propto n_z^2-(n_x^2+n_y^2)/2}$ is the leading anisotropy which will drive the transition, and $X^{(4)}_{3333}$ is a subleading anisotropy.
The scaling dimension formula Eq.~\ref{eq:xk} is reliable only at large $S$, but it suggests  that for small spin there is a range of positive $\delta$ where
 $X^{(2)}$ is the only relevant anisotropy, $X^{(4)}$ and higher anisotropies being irrelevant.  We assume $\delta$ is in this range. 
 
 Then, the $\mathrm{SO}(3)$-invariant fixed point governs a phase transition line
in the $(\Delta/J, \gamma)$ plane. One point on this line, at $(0, 1)$, has microscopic $\mathrm{SO}(3)$ symmetry, but at other points on the line $\mathrm{SO}(3)$ emerges only in the IR. 
One adjacent phase is the easy-axis phase, where $\mathbb{Z}_2$ is broken.
The nature of the other phase will depend on the spin. For  spin-1/2 it is likely a power-law phase \cite{sengupta2000spin} in which the the easy-plane order parameter dominates.

It may be interesting to check for symmetry enhancement starting from other microscopic symmetry groups. For example $S_4$ (tetrahedral) symmetry allows the perturbation $X^{(3)}_{123}$. 
We may argue that the field theory with this symmetry breaking, and with $S=1$, is an effective theory for a long-range 4-state Potts model in which the partition sum is weighted by $(-1)$ for each domain wall.

\newsec{Conclusions.} The impurity model can be seen as the simplest member of a dimensional hierarchy of sigma models with topological terms \cite{AbanovWiegmann}.
We have argued that some interesting features of the RG flows in higher dimensions are also present in 0+1D, giving a rich phase diagram for the Bose-Kondo model.
The model yields an example of fixed point annihilation 
that is tractable both analytically and in simulations,
and also shows  analogues of phenomena from higher-dimensional ``non-Landau'' phase transitions.
It would be interesting to examine other variations --- for example models in large $N$ limits, with other symmetric spaces for the target space, or with coupling to fermions --- and to explore physical realizations of the tunable interaction exponent $\delta$ (perhaps via bosonic bath  whose hopping parameters varied with distance from the impurity).
Finally it would also be interesting to look for the annihilation phenomenon in models
relevant to impurities in critical magnets  in which  the bath is not Gaussian  \cite{sachdev1999quantum},
or settings where the impurity arises as a self-consistent description of an interacting many-body system \cite{chowdhury2021sachdev}.

\newsec{Related work:} The impurity in the large $S$ limit has also been analyzed recently in two other papers, Ref.~\cite{cuomo2022spin} and Ref.~\cite{beccaria2022wilson}, with results for the beta function consistent with those above. In addition, these papers   make interesting connections with Wilson lines and line defects in  conformal field theory. Quantum Monte Carlo results for spin-1/2 are now also available \cite{weber20222}, and are consistent with the phase diagram obtained here.

\newsec{Acknowledgements:} I am grateful to D. Bernard, X. Cao, M. Metlitski, and T. Senthil for useful discussions, and to D.~Bernard also for comments on the draft. I thank the authors of Ref.~\cite{cuomo2022spin} for correspondence. I acknowledge support from a Royal Society University Research Fellowship during part of this work.

\appendix

\section{RG calculation}
\label{app:calculation}

We integrate out fast fluctuations of the field $\vec n(t)$ around a slowly varying background $\vec n_s(t)$.
In order to determine the flow of the coupling, it is sufficient to take $\vec n_s(t)$ to lie in the XY plane,
\be
\vec n_s(t) = \lf \cos \phi_s(t), \sin\phi_s(t), 0 \ri.
\ee
Here we are exploiting the fact that the renormalization of the coupling in a given RG step is independent of the slow field configuration \cite{polyakov1987gauge}.
We can also take $\vec n_s(t)$ to be a state of uniform twist, $\ddot \phi_s=0$, which simplifies the expansion in the fast modes slightly.
We may parameterize these with fields $\phi_f$, $\chi_f$, with frequencies $|\omega|$ in the range ${[\Lambda/b, \Lambda]}$, with ${b=e^{\Delta \tau}}$, as
\be
\vec n = \lf 
\sqrt{1-\chi_f^2}
 \lf \cos (\phi_s+\phi_f),
 \sin (\phi_s+\phi_f) \ri, 
 \chi_f 
\ri.
\ee
We then expand the action to quadratic order in the fast modes. 
Taking the state $\vec n_s$ of uniform twist, the linear terms in $\phi_f$, $\chi_f$ vanish and
\ba
\Omega[\vec n] & \simeq \Omega[\vec n_s]  - \int \dd t \chi_f \partial_t \phi_f,
\\
\notag (\vec n - \vec n' )^2 & \simeq
 (\chi_f - \chi_f')^2 + (\phi_f - \phi_f')^2 \\
& + (\vec n_s - \vec n'_s )^2  \Big(
1 - \f{1}{2} \left[ \chi_f^2 + \chi_f'^2 +(\phi_f-\phi_f')^2 \right]
\Big).  \notag
\end{align}
Let us rescale ${(\phi_f, \chi_f) \rightarrow (\phi_f, \chi_f)/\sqrt{S}}$. 
We also drop the subscripts on the fields, since from now on it is implied  that $\vec n=\vec n_s$ and $(\phi,\chi) = (\phi_f, \chi_f)$.
Suppressing the integral measures and the time arguments, and writing $\vec n$ and $\vec n'$ for $\vec n(t)$ and $\vec n(t')$ etc., the action is
\ba
\mathcal{S} = \mathcal{S}_f + \mathcal{S}_{fs}
\end{align}
with
\ba
\mathcal{S}_f = \iint   \f{K}{2h}\left[ (\phi-\phi')^2 + (\chi-\chi')^2\right] + i \int \dd t \, \chi\dot \phi
\end{align}
and
\ba
\mathcal{S}_{fs} =  S\lf  \iint \f{K\times (1-R)}{2h} (\vec n-\vec n')^2   - i \Omega[\vec n] \ri.
\end{align}
The term $R$ is
\ba
R =  \f{1}{2S} \left[ \chi^2 + \chi'^2 +(\phi-\phi')^2 \right]. 
\end{align}
We may integrate out the fast fields using a cumulant expansion (see e.g. \cite{cardy1996scaling})
\be
\<  e^{-\mathcal{S}_{fs}} \> = e^{- \< \mathcal{S}_{fs}\>  + \ldots },
\ee
where the average is taken using the action $\mathcal{S}_f$.
The simplification is that at large $S$ we can keep only the leading term in the cumulant expansion, because the part of $\mathcal{S}_{fs}$ that depends on the fast fields is of order $1/S$.\footnote{The leading part of the long-range coupling in $\mathcal{S}_{fs}$, 
of the form ${S\times K(n-n')^2}$, is of order $S^0$ (in order to avoid exponential suppression of the Boltzmann weight), 
so the nontrivial correction term in the exponent, coming from $R$, is of order $1/S$.}
The required average is 
\ba\notag
\< R\> & = \f{1}{S} \left[  \< \chi^2\> + \<\phi^2\> - \< \phi \phi'\> \right] \rightarrow \f{1}{S} \left[  \< \chi^2\> + \<\phi^2\>  \right].
\end{align}
We have dropped the two-point function of the fast field $\phi$ at distinct points: since this is oscillatory and decaying, it does not contribute to renormalizing the coupling of the power law interaction.

Therefore integrating out the fast fields effects the change
$1/h \rightarrow (1 - \< R\>)/h$.
After rescaling our temporal coordinate  in order to restore the UV cutoff to the initial value $\Lambda$ we have the RG transformation
\be\label{eq:rgstep}
h(\tau+\Delta\tau) \simeq e^{-\delta \Delta \tau} h(\tau) (1 + \< R\>).
\ee

It remains to find $\<\phi^2\>$ and $\<\chi^2\>$.
With the normalisation of $K(t)$ in Eq.~\ref{eq:Kdefn} of the main text (where $\Gamma$ is the Gamma function)
the Fourier transform satisfies ${2(\widetilde K(0)-\widetilde K(\omega))=\Lambda^\delta |\omega|^{1-\delta}}$, and in frequency space
\be\notag
\mathcal{S}_f = 
\f{1}{2} \int \f{\dd \omega}{2\pi} 
\Psi(\omega)^T
\left(
\begin{array}{cc}
\gamma^{-1} |\omega|^{1-\delta}  &  \omega     \\
 - \omega &      \gamma^{-1} |\omega|^{1-\delta}
\end{array}
\right) \Psi(-\omega)
\ee
with ${\Psi = (\chi, \phi)^T}$ and ${\gamma=\Lambda^{-\delta}h}$, so that 
\ba\notag
\<\Psi_i(\omega) \Psi_j (\omega')\> = &  
\f{2\pi  \delta(\omega+ \omega') }{
|\omega|^{1-\delta} (1+ \gamma^2 |\omega|^{2\delta})
}\\ \notag
& \quad \times
\left(
\begin{array}{cc}
\gamma &  \gamma^2 \omega |\omega|^{\delta-1} \\
 -\gamma^2 \omega |\omega|^{\delta-1}  &  \gamma
\end{array}
\right)_{ij}.
\end{align}
Therefore
\ba
\<\chi^2\> = \< \phi^2\> & = 
 2 \int_{\Lambda/b}^{\Lambda} \f{\dd \omega}{2\pi} 
\f{1}{\omega^{1-\delta}} \f{\gamma}{1+ \gamma^2 \omega^{2\delta} } \\
& =
 2 h   \int_{e^{-\Delta \tau}}^{1} \f{\dd v}{2\pi} 
\f{1}{v^{1-\delta}} \f{1}{1+ h^2 v^{2\delta} }.
\end{align}
We are free to take any value of $\Delta \tau$ so long as ${\Delta \tau/S \ll 1}$ \cite{polyakov1975interaction}, but it is simplest to see the result if we take $\Delta \tau$ infinitesimal, in which case
\be
\<\chi^2\> = \< \phi^2\> = \f{\Delta \tau}{\pi} \f{h}{1+h^2}.
\ee
Therefore from (\ref{eq:rgstep}) 
\be\label{eq:rgeqapp}
\f{\dd h}{\dd \tau} =  - \delta h + \f{2}{\pi S} \f{h^2}{1+h^2},
\ee
as stated in the main text.

Next, consider the renormalization of operators.  The running scaling dimension $x_1(h)$ of $\vec n$ is determined by~\cite{polyakov1975interaction}
\be
\< \vec n(t) \>_\text{fast} \simeq e^{- x_1(h) \Delta \tau} \, \vec n_s(t).
\ee
It is convenient to take ${\vec n_s=(1,0,0)}$, so that ${\vec n\simeq\lf 1-\f{1}{2S}\< \chi^2 \> -\f{1}{2S} \<\phi^2\>,0,0\ri}$, and
\be\label{eq:x1h}
x_1(h) = \f{1}{\pi S} \f{h}{1+h^2}.
\ee
At a nontrivial fixed point ({$h\neq 0,\infty$}) this is $x_1=\delta/2$, because (\ref{eq:rgeqapp}) has the form ${\dot h = ( - \delta  + 2 x(h))h}$.

The operator $n$ transforms in the spin-1 representation of $\mathrm{SO}(3)$. We can make spin-$k$ operators ${X^{(k)}_{a_1,\ldots, a_k} = n_{a_1}\ldots n_{a_k} - (\ldots)}$,
where the $(\ldots)$ represents terms subtracted to make $X^{(k)}$ traceless.
The ratio between the running scaling dimension $x_k$ of $X^{(k)}$ and that of $n$ is the same as in the $\mathrm{O}(3)$ nonlinear sigma model in ${2+\epsilon}$ dimensions~\cite{brezin1976renormalization}:
\be
x_k(h) = \f{k(k+1)}{2} x_1(h).
\ee
For example, consider the renormalization of ${X^{(2)}_{ab}=n_an_b-\delta_{ab}/3}$. It is sufficient to consider a single component to extract the scaling dimension:
\be\label{eq:Xfastav}
\< X^{(2)}_{11}\>_\text{fast} = e^{- x_2(h) \Delta \tau} (X^{(2)}_s)_{11}.
\ee
Again take $n_s = (1,0,0)$. Then the equation above is
\be\label{eq:n1sq}
\<n_1^2\>_\text{fast} - \f{1}{3} = \lf 1-x_2(h) \Delta \tau \ri  \f{2}{3} .
\ee
Since (from the expansion of $\vec n$ above Eq.~\ref{eq:x1h})
\be\label{eq:n1k}
\<n_1^k\>_\text{fast} = 1 - k \, x_1(h) \Delta \tau,
\ee
Eq.~\ref{eq:n1sq} gives $x_2(h) = 3 x_1(h)$. We can proceed similarly for general $k$. The combinatorial factors depend only on group theory so they are  the same as for the model in $2+\epsilon$ dimensions~\cite{brezin1976renormalization}.
 The result is valid in the limit where $S\gg k$ (see Eq.~\ref{eq:n1k}, where the second term, which was assumed small, is of order $k/S$). It also would be interesting to examine the separate regime $k\sim S$.

Finally we discuss correlators $G^{(k)}(t)$ of higher-spin operators  with ${k>2S}$. 
In the text we suggested that the ${t\rightarrow\infty}$ limit of such corrrelators distinguished  the free spin phase (which exists at  $\delta \leq 0$) from the ordered phase which exists for $\delta >0$.

First let us clarify the  meaning of such operators in the microscopic theory of a spin coupled to a bath. 
For a free spin-$S$ (without a coupling to a bath), nonvanishing operators exist only in representations of $\mathrm{SO}(3)$ with spin ${k\leq 2S}$. Taking the case  $S=1/2$ as an example, we have Pauli operators $\hat \sigma_a$ which transform in the spin-1 representation, but no spin-2 operators. 

However, in the model of a spin coupled to a local bath magnetization $\hat m_a$, operators with arbitrary spin can be written down.
For example, ${\hat O_{ab} = {\f{1}{2} ( \hat m_a  \hat \sigma_b + \hat m_b \hat \sigma_a)  - \f{1}{3}  \delta_{ab} \hat m. \hat \sigma}}$ 
transforms in the spin-2 representation and is nonzero
(in contrast to the analogous operator with  $\hat m$  replaced by $\hat \sigma$, which vanishes by the Pauli anticommutation relations).

Alternately, we can write higher spin operators entirely in terms of the spin degree of freedom, by using products of Heisenberg picture operators $\hat \sigma_a(t)$  at distinct but nearby times. 
For fixed $\Delta t$, an operator such as (we make the time argument explicit)
\be\label{eq:Oprime}
\hat O'_{ab}(0) \equiv \f{ \hat \sigma_a(\Delta t)  \hat \sigma_b(0) +  \hat \sigma_b(\Delta t)  \hat \sigma_a(0)}{2}  - \f{\delta_{ab} \hat \sigma(\Delta t). \hat \sigma(0)}{3}  
\ee
is still a local operator from the point of the RG.  This operator  has the same symmetry as $\hat O_{ab}$ above, so we expect it to coarse-grain to  the same continuum operator.  
This can be seen easily at small $\Delta t$, when  ${\hat  \sigma_a(\Delta t)\simeq e^{\Delta t \hat H_\text{int}(0)} \hat\sigma_a(0) e^{- \Delta t \hat H_\text{int}(0)}}$ is approximately  ${\hat  \sigma_a(\Delta t) \simeq 
\hat \sigma_a(0) + i J \Delta t \, \epsilon_{abc} \,\hat \sigma_b(0) \hat m_c(0)}$. Substituting into (\ref{eq:Oprime}) gives ${\hat O'_{ab}(0) \simeq - J \Delta t \, \hat O_{ab}(0)}$. Physically, the need to separate the two $\hat \sigma$ insertions in time is because the spin is only able to absorb 2 successive units of angular momentum if it has time to exchange angular momentum with the bath in between.

On grounds of symmetry, we identify these microscopic  spin-$k$ operators (up to normalization, and subleading terms) with the operators  $X^{(k)}$ in the coarse-grained theory, e.g. ${X^{(2)}_{ab} = n_a n_b - \f{1}{3} \delta_{ab} n^2}$. Below we use $G^{(k)}(t)$ to represent the two-point function of any nonvanishing microscopic operator with the right symmetry.

In the ordered phase (at $\delta>0$), $G^{(k)}(t\rightarrow \infty)$ will be nonzero for all $k$, with a magnitude that is reduced by fluctuations (since the bare value of $h$ is nonzero), cf. Eq.~\ref{eq:Xfastav}.

In the free spin phase for $\delta \leq 0$, 
$h$ flows to infinity and the spin decouples from the bath in the IR.
In this phase, $G^{(k)}(\infty) = 0$ for ${k>2S}$.
The presence of the bath means that  $G^{(k)}(t)$ is not strictly zero for finite $t$.
For simplicity, consider ${\delta<0}$ (for ${\delta=0}$ there are logarithmic corrections): then we expect  $G^{(k)}(t)\sim t^{-(2-\delta)(k-2S)}$ at large $t$.
The corresponding spin-$k$  operator in the decoupled spin/bath fixed point theory is made from a spin-$2S$ operator acting on the spin (whose autocorrelator is time-independent), and a spin ${k-2S}$ operator  ${(\hat m_{a_1}\ldots \hat m_{a_{k-2S}}-\ldots)}$  whose autocorrelator is given by Wick's theorem.

\bibliography{impurityrefs}
\end{document}